\documentclass[usenatbib,twocolumn,8pt]{article}

\usepackage{amssymb}
\usepackage[fleqn]{amsmath}
\usepackage{times}
\usepackage{graphicx}
\usepackage{subfigure}
\usepackage[T1]{fontenc}
\usepackage{aecompl} 
\usepackage{multirow}
\usepackage{pdflscape}
\usepackage{rotating}

\usepackage{mathrsfs}
\usepackage{xcolor}

\usepackage{anyfontsize}

\usepackage[breaklinks=true]{hyperref}
\hypersetup{colorlinks=true,citecolor = blue}

\usepackage{verbatim}
\usepackage[numbers]{natbib}
\usepackage{multibib}
\newcites{body}{References}
\usepackage{natbib}
\newcommand{\mnras}{MNRAS}

\begin{document}

\baselineskip=5mm

\twocolumn[{

\noindent {\large {\color{blue}\textrm{Note}}}\\
\noindent\rule{19cm}{0.4pt}\\

{\large{\bf  The paper `How proper are Bayesian models in the astronomical literature?' [arxiv:1712.03549] by Tak, Ghosh and Ellis is improper \\}} 

{\normalsize \bf Mauro Sereno}, mauro.sereno@oabo.inaf.it \\

{\footnotesize
\noindent 
}

}]

\smallskip

\small

\noindent \textbf{
In their `How proper are Bayesian models in the astronomical literature?' \citep[arXiv:1712.03549]{tak+al17}, Hyungsuk Tak, Sujit K. Ghosh and Justin A. Ellis criticised my work with false statements. This is an infamous case of straw man fallacy. They give the impression of refuting an opponent's argument, while they refute an argument that was not presented.}


\noindent

Hyungsuk Tak, Sujit K. Ghosh and Justin A. Ellis came to our rescue. Astronomers need Tak, Ghosh and Ellis, and Tak, Ghosh and Ellis are willing to patrol. Tak, Ghosh and Ellis put themselves in charge. 

Tak, Ghosh and Ellis decided to grade other researchers. I am one of them. Unfortunately, they utterly misquoted my work.  Apparently, their reasoning is solid. Their premises are true. Still, their argument is viciously fallacious. In their abstract, they wrote `the posterior distribution may no longer be a probability distribution if an improper prior distribution (non-probability measure) such as an unbounded uniform prior is used'. We can all agree on this.

Unfortunately, they improperly attributed the use of the unbounded uniform prior to some astronomers. This is unfair. They also decided that I was one of them. I am not.

In the prestigious Tak-Ghosh-Ellis system, a paper of mine is ranked a (b). On pag. 7, they wrote `Three articles published online in MNRAS employ improper priors without proving posterior propriety; Sereno and Ettori (2017) use an improper uniform prior on $\texttt{mu.Z.0}$ whose upper limit is infinity ($\texttt{Z.max} = +\infty$ in Table 1)'.  This is not correct.

As clearly reported in table 1 of \citet[2017,][]{se+et17_comalit_V}, the default prior for $\texttt{mu.Z.0}$ is $\texttt{dunif}$, which is constant in a bounded range and null otherwise. $\texttt{Z.max}$ is the maximum value of $Z$, a latent variable not to be confused with the parameter $\texttt{mu.Z.0}$. If not customized otherwise, $-\texttt{n.large} <  \texttt{mu.Z.0}< \texttt{n.large}$, with $\texttt{n.large}=10^4$, whereas $Z$ follows a Gaussian distribution with mean equal to $ \langle Z \rangle = \texttt{mu.Z.0}$. The Gaussian distribution is obviously unbounded (then $\texttt{Z.max} \rightarrow +\infty$). The prior on $\texttt{mu.Z.0}$ was also reported in \citep[table~1]{ser16_lira}, \citep[eq.~(12)]{se+et15_comalit_I}, \citep[eq.~(9)]{ser+al15_comalit_II} and \citep[eq.~(28)]{se+et15_comalit_IV}.

All the details are given in \citet[2017,][]{se+et17_comalit_V} and references therein, mainly the manual of the employed software \textsc{LIRA}\footnote{The \textsc{R}-package \textsc{LIRA} (LInear Regression in Astronomy) is publicly available from CRAN (Comprehensive R Archive Network) at \url{https://cran.r-project.org/web/packages/lira/index.html}.}, the accompanying LIRA paper \citep{ser16_lira}, and the CoMaLit series of papers \citep[COmparing MAsses in LITerature, ][]{se+et15_comalit_I,ser+al15_comalit_II,ser15_comalit_III,se+et15_comalit_IV,se+et17_comalit_V}.

\textsc{LIRA} exploits the \textsc{JAGS} libraries\footnote{\textsc{JAGS} (Just Another Gibbs sampler) by M.~Plummer performs analysis of Bayesian hierarchical models. It is publicly available and fully referenced at \url{http://mcmc-jags.sourceforge.net/}.} and warns against improper priors. The software is free and open-source. The code \citep[app. A,][]{se+et17_comalit_V} and the data-sets\footnote{The Literature Catalogs of weak Lensing Clusters (LC$^2$) can be downloaded from \url{http://pico.oabo.inaf.it/\textasciitilde sereno/CoMaLit/LC2/}. See \citep{ser15_comalit_III}.} used in \citep{se+et17_comalit_V} were made publicly available. As an output, \textsc{LIRA} can export the hierarchical model too. The interested reader can check and reproduce all the steps of the analysis. 

The work of other astronomers `classified' in the Tak-Ghosh-Ellis `categories' may be misquoted too. I asked privately the authors of \citep{tak+al17} to rectify. They refused, even though they keep updating their post.
 
The use of informal fallacies in the scientific conversation is spreading out. This note is an attempt to unmask one such abuse.

{\footnotesize
\renewcommand{\refname}{\small References}

}

\end{document}